\documentclass[jkps,twocolumn,fleqn,showpacs,showkeys,superscriptaddress]{revtex4-1}
\usepackage[pdftex]{graphicx}
\usepackage{amssymb}
\usepackage{amsmath}
\usepackage{bm}
\usepackage{multirow}
\usepackage{xcolor}
\usepackage{soul}
\usepackage{upgreek}

\begin{document}
\title[]{Simulation of Classical Axion Electrodynamics using COMSOL Multiphysics}
\author{Junu \surname{Jeong}}
 \email{jwpc0120@ibs.re.kr}
 \affiliation{Center for Axion and Precision Physics Research, Institute for Basic Science (IBS), Daejeon 34051, Republic of Korea}
\author{Younggeun \surname{Kim}}
 \affiliation{Center for Axion and Precision Physics Research, Institute for Basic Science (IBS), Daejeon 34051, Republic of Korea}
\author{Sungjae \surname{Bae}}
 \affiliation{Department of Physics, Korea Advanced Institute of Science and Technology (KAIST), Daejeon 34141, Republic of Korea}
 \affiliation{Center for Axion and Precision Physics Research, Institute for Basic Science (IBS), Daejeon 34051, Republic of Korea}
\author{Sungwoo \surname{Youn}}
 \affiliation{Center for Axion and Precision Physics Research, Institute for Basic Science (IBS), Daejeon 34051, Republic of Korea}

\date{\today}

\begin{abstract}
The axion is a hypothetical particle motivated to address the strong CP problem, and is one of the appealing dark matter candidates.
Numerous experimental searches for dark matter axions have been proposed relying on their coupling with photons.
The classical equations of motion for the axion-photon coupling are well known but need to be fully computed for complex experimental setups.
The partial differential equations of axion electrodynamics can be numerically solved using finite element methods.
In this work, we simulate axion electrodynamics using \texttt{COMSOL Multiphyics}, a commercially available simulation software, for various experimental schemes, including the dish antenna haloscope, cavity haloscope, dielectric haloscope, and axion-photon regeneration.
We show that the numerical results are in good agreement with the analytical solutions.
\end{abstract}


\maketitle

\section{\label{sec:intro}Introduction}
The axion is a pseudo-Goldstone boson resulting from the Peccei-Quinn symmetry, which was proposed to solve the CP problem in quantum chromodynamics (QCD)~\cite{PRL1977PQ,PRL1978Weinberg,PRL1978Wilczek}.
It is also one of the leading candidates for cold dark matter, if its mass lies between $\mathcal{O}\left(\mu {\rm eV}\right)$ and $\mathcal{O}\left( {\rm meV}\right)$~\cite{PLB1983Wilczek,PLB1983Abbott,PLB1983Dine}.
The invisible axions can interact weakly with the Standard Model particles depending on theoretical models, notably KSVZ~\cite{PRL1979Kim,NPB1980SVZ} and DFSZ~\cite{YF1980Zhitnitsky,PLB1981DFS}.
The Lagrangian density describing the axion-photon interaction, which many experimental searches rely on, is given by~\cite{PRL1987Wilczek}.
\begin{equation}
    \label{eq:aEB}
    \mathcal{L}_{a\gamma\gamma} = -\frac{g_{a\gamma\gamma}}{Z_{0}} a \mathbf{E} \cdot \mathbf{B},
\end{equation}
where $g_{a\gamma\gamma}$ is the axion-photon coupling in the unit of ${\rm GeV}^{-1}$, $Z_{0}=\sqrt{\mu_{0}/\varepsilon_{0}}$ is the impedance of free space, $a$ is the axion field in the unit of ${\rm GeV}$, and $\mathbf{E}$ and $\mathbf{B}$ are the electric and magnetic fields, respectively.
A number of experimental ideas have been proposed to probe the feeble interaction of axion dark matter with photons using relatively small-scale setups.
One of the most sensitive approaches is the haloscope, a technique designed to observe the dark matter halo utilizing cavities~\cite{PRD1985Sikivie}, dielectric arrays~\cite{PRL2017MADMAX}, or dish antennae~\cite{JCAP2013Horns}.
There is also an independent search scheme where axions can be generated and detected in the laboratory without requiring astrophysical or cosmological sources~\cite{PRL1987Bibber}.
This offers the basis for so-called light-shinning-through-a-wall experiments.

Due to the complex geometry of the setup and the inhomogeneity of some parameters, experimental sensitivities usually require numerical calculations.
One of the leading simulation tools is \texttt{COMSOL Multiphysics}~\cite{COMSOL}, a commercially available software for solving partial differential equations (PDE) based on the finite element method (FEM).
In particular, the {\it RF (Radio Frequency) module} included in the package provides the ability to design and optimize RF devices and components by solving Maxwell's equations for arbitrary geometries.
This feature enables us to numerically calculate the classical axion-photon interactions for various haloscopes.
The photon regeneration scheme can also be simulated by implementing the Klein-Gordon equation for pseudo-scalar particles.

In this work, we show how to use the simulation package to compute classical axion electrodynamics numerically for various experimental schemes.
For verification, we compare the results with those obtained analytically.
Section~\ref{sec:axionEM} reviews the properties of dark matter axions and classical electrodynamics including their interactions with photons.
Section~\ref{sec:COMSOL} gives a brief instruction on how to implement axion electrodynamics in \texttt{COMSOL} software.
In Section~\ref{sec:application}, various haloscope setups are considered to test the numerical simulation and the results are compared with the analytic solutions.
Finally, the photon-regeneration scheme is also modeled by solving the Klein-Gordon equation.

\section{\label{sec:axionEM}Axion Electrodynamics}
Taking the axion-photon coupling in Eq.~\ref{eq:aEB} into account, Maxwell's equations are modified as~\cite{PRL1987Wilczek}
\begin{equation}
\begin{split}
    \varepsilon \nabla \cdot \mathbf{E} & = \rho_{o} + \rho_{a}, \\
    \nabla \cdot \mathbf{B} & = 0, \\
    \nabla \times \mathbf{E} + \dot{\mathbf{B}} & = 0 , \\
    \mu^{-1}\nabla \times \mathbf{B} - \varepsilon \dot{\mathbf{E}} & = \mathbf{J}_{o} + \mathbf{J}_{a}. \\
\end{split}
\label{eq:mod_ME}
\end{equation}
Here $\varepsilon$ and $\mu$ are the permittivity and permeability, and $\rho_{o}$ and $\mathbf{J}_{o}$ are the ordinary charge and current densities, respectively.
$\rho_{a}$ and $\mathbf{J}_{a}$ are the axion induced charge and current densities with the form of
\begin{equation}
\begin{split}
    & \rho_{a} = \frac{g_{a\gamma\gamma}}{Z_{0}} \left( \nabla a \right) \cdot \mathbf{B}, \\
    & \mathbf{J}_{a} = -\frac{g_{a\gamma\gamma}}{Z_{0}} \left( \dot{a}\mathbf{B} + \left(\nabla a\right) \times \mathbf{E} \right), \\
\end{split}
\label{eq:axion-cc}
\end{equation}
and they naturally satisfy the charge conservation law: $\dot{\rho_{a}}+\nabla\cdot\mathbf{J}_{a} = 0$.

For dark matter axions, the gradient terms ($\nabla a$) in Eq.~\ref{eq:axion-cc} are usually ignored by the definition of `cold' (non-relativistic).
Therefore, an axion field can be approximated as $a \approx a_{0}e^{-i \omega_{a} t}$, where $a_{0}$ is the amplitude and $\omega_{a}$ is the angular frequency of the field.
The field amplitude is related to the local dark matter density ($\varrho_{a}$) as:
\begin{equation}
    a_{0} = \frac{\sqrt{2 \varrho_{a} \hbar c^{3}}}{\omega_{a}},
\label{eq:a0}
\end{equation}
with $\varrho_{a} = 0.3 \textrm{--} 0.45\,{\rm GeV/cm^{3}}$ in the solar system~\cite{JPG2014Read}.
The angular frequency determines the axion mass ($m_a$) by $\omega_{a} \approx m_{a} c^{2} / \hbar$, where $c$ is the speed of light and $\hbar$ is the reduced Planck constant.
For axion-photon interactions, in particular, the coupling constant is given by
\begin{equation}
    g_{a\gamma\gamma} = \left(g_{\gamma} \frac{\alpha}{\pi} \right) \frac{m_{a}c^{2}}{\sqrt{\chi}},
\label{eq:g_arr}
\end{equation}
where $g_{\gamma}$ is the model dependent coefficient, $\alpha$ is the fine structure constant, and $\chi$ is the zero-temperature topological susceptibility of QCD that is expected to be around $\left(75.6\,{\rm MeV}\right)^{4}$~\cite{NATURE2017Borsanyi}.
The representative invisible axion models, KSVZ and DFSZ, have $g_{\gamma}$ values of $0.97$ and $-0.36$, respectively.
The product of Eqs.~\ref{eq:a0} and~\ref{eq:g_arr} is a dimensionless quantity:
\begin{equation}
\begin{split}
    g_{a\gamma\gamma}a_{0} &= \left(g_{\gamma}\frac{\alpha}{\pi} \right) \sqrt{\frac{2 \varrho_{a} \hbar^{3} c^{3}}{\chi} } \\
    &\approx 10^{-21} \left(\frac{g_{\gamma}}{0.97} \right) \sqrt{\frac{\varrho_{a}}{0.45\,{\rm GeV/cm^{3}}} \frac{(75.6\,{\rm MeV})^{4}}{\chi} }.
\end{split}
\end{equation}

Since $g_{a\gamma\gamma} a_{0} \ll 1$ for invisible axions, the equations associated only with the reacted field, decoupled from modified Maxwell's equations (Eq.~\ref{eq:mod_ME}), can be approximated as~\cite{PDU2019Kim}
\begin{equation}
\label{eq:axion_haloscope_maxwell}
\begin{split}
    \varepsilon \nabla \cdot \mathbf{E}_{r} & = \frac{g_{a\gamma\gamma}}{Z_{0}} \left( \nabla a \right) \cdot \mathbf{B}_{o} \approx 0, \\
    \nabla \cdot \mathbf{B}_{r} & = 0, \\
    \nabla \times \mathbf{E}_{r} + \dot{\mathbf{B}}_{r} & = 0, \\
    \mu^{-1} \nabla \times \mathbf{B}_{r} - \varepsilon \dot{\mathbf{E}}_{r}
        & = - \frac{g_{a\gamma\gamma}}{Z_{0}} \left( \dot{a}\mathbf{B}_{o} + \left(\nabla a\right) \times \mathbf{E}_{o} \right) \\
        & \approx \frac{i\omega_{a}}{Z_{0}} \left( g_{a\gamma\gamma} a_{0} e^{-i\omega_{a} t} \right) \mathbf{B}_{o} , \\
    \end{split}
\end{equation}
where $\mathbf{E}_{o}$ and $\mathbf{B}_{o}$ are the ordinary electric and magnetic fields that satisfy Maxwell's equations, and $\mathbf{E}_{r}$ and $\mathbf{B}_{r}$ are the first-order electric and magnetic fields reacted by dark matter axions.
The last equation of Eq~\ref{eq:axion_haloscope_maxwell} states that the oscillating current density induced by dark matter axions can drive an additional electromagnetic field proportional to the ordinary magnetic field.

\section{\label{sec:COMSOL}Simulation method}
Most haloscope experiments attempt to detect reacted electromagnetic fields under a static magnetic field.
In this case, the frequency of the reacted fields becomes that of the axion.
For coupled linear differential equations consisting of a single frequency component, the frequency response can be obtained by Fourier transformation.

\texttt{COMSOL}, one of the representative commercially available PDE solvers based on FEM, facilitates various physical phenomena in complex geometry with a user-friendly interface~\cite{COMSOL}.
In particular, {\it RF module} in \texttt{COMSOL} solves Maxwell's equations for an arbitrary geometry and boundary conditions.
The physics of {\it Electromagnetic Waves, Frequency Domain} in {\it RF module} accurately simulates the solution of electromagnetic waves in the frequency domain.
It also supports adding arbitrary current densities in the frequency domain study with the feature of {\it External Current Density}.
Users can simulate the reacted electromagnetic fields by substituting $\mathbf{J}_{e} = i g_{a\gamma\gamma}a_{0}\omega_{a} \mathbf{B}_{o}/Z_{0}$ to the node of {\it External Current Density} according to the direction of the ordinary magnetic field.
Recently, a study of the same method was conducted in the DMRadio collaboration for lumped element experiments~\cite{arxiv2023DMRadio}.

Once the reacted field solution is obtained, the conversion power can be calculated from the Ohmic-like loss by the axion inducing current density~\cite{PDU2019Kim}.
\begin{equation}
    P_{a\gamma\gamma} = -\frac{1}{2} \Re \left[ \int \mathbf{J}_{a}^{*}\cdot \mathbf{E}_{r} dV \right],
\end{equation}
where $\Re$ refers to the real operator and $V$ is the volume of the simulated domain.
From the energy conservation law, the conversion power can also be estimated as a sum of the power loss:
\begin{equation}
\begin{split}
    P_{a\gamma\gamma} (= P_{\rm loss}) & = \frac{1}{2} \int R_{s} \left|\frac{1}{\mu}\mathbf{B}_{r}\times\hat{n} \right|^{2} dS \\
    & + \frac{1}{2} \int \omega_{a}\varepsilon \tan \delta  \left| \mathbf{E}_{r} \right|^{2} dV \\
    & + \frac{1}{2} \Re\left[ \int \frac{1}{\mu} \left( \mathbf{E}_{r} \times \mathbf{B}_{r}^{*} \right) \cdot \hat{n} \right] dS,
\end{split}
\end{equation}
where $R_{s}$ is the surface resistance of the boundary conductor, $\tan \delta$ is the tangent delta of dielectrics, $S$ is the surface of the integration boundary, and $\hat{n}$ is the normal vector to the surface.
The first term is the loss from the conductor boundary, the second term is the loss from the dielectrics volume, and the last term is the radiation loss.

\section{\label{sec:application}Axion haloscope}
In this section, representative haloscope experiments of searches for dark matter axions are introduced.
The reacted electromagnetic fields in a simplified geometry are simulated and the conversion powers are calculated.
In each case, we confirm the simulation results by comparing them with those from the analytic solution.

\subsection{\label{subsec:dah}Dish antenna haloscope}
According to Eq.~\ref{eq:axion_haloscope_maxwell}, dark matter axions on a metal plate under a magnetic field are converted into photons.
The converted photons are emitted in a direction perpendicular to the metal plate.
The dish antenna haloscope gathers the radiated photons into one point with a spherical or parabolic mirror and measures them with a photon detector.
Since enhancement through resonance is not used, axion masses in a wide range can be searched without frequency tuning.

The simplest version of the dish antenna haloscope is an infinite metal plate.
Assuming that the magnetic field of $B_{0}$ is applied from $x=0$ to $x=L$ in the $y$-axis direction, and the metal plate is placed perpendicular to the $x$-axis at $x=0$, the reacted radiating electric field and the corresponding Poynting vector at $x > L$~\cite{JCAP2013Horns} are:
\begin{equation}
\begin{split}
    & \mathbf{E}_{r}(x > L) = -i c g_{a\gamma\gamma}a_{0} B_{0} (\cos(k L) - 1) e^{i k x} \hat{y}, \\
    & \mathbf{S}_{r} = \frac{1}{2 \mu} \mathbf{E}_{r} \times \mathbf{B}_{r}^{*} = \frac{\left|g_{a\gamma\gamma}a_{0} \right|^{2}}{2} \frac{c}{\mu}B_{0}^{2} \left(4 \sin^{4} \left(\frac{k L}{2}\right) \right) \hat{x}.
\end{split}
\end{equation}
Here $k=\omega_{a}/c$ is the wave-number.
Since the magnetic field component perpendicular to the metal plate does not affect the boundary condition, we can write the radiation power without loss of generalization as:
\begin{equation}
\label{eq:DAH_power}
\begin{split}
    \frac{P_{a\gamma\gamma}}{A} &=
    \Re\left[\mathbf{S}_{r}\cdot \hat{n} \right] \\
    &= \left[ \frac{g_{a\gamma\gamma}^{2}\varrho_{a}}{m_{a}^{2}c /\hbar^{3}} \right] \frac{c}{\mu}|\mathbf{B}_{o} \times \hat{n}|^{2} \left( 4 \sin^{4} \left(\frac{\omega_{a} L}{2 c}\right) \right),
\end{split}
\end{equation}
where $A$ is the area of the metal surface and $\hat{n}$ is the normal vector to the metal plate.
The term in square brackets is the model-dependent parameter with no units ($=\left|g_{a\gamma\gamma}a_{0}\right|^{2}/2$).
The sine-to-the-power-of-4 in the last parentheses has an average of 1.5, and a median of 1 for frequencies.

In \texttt{COMSOL}, an infinite dish antenna can be effectively simulated through periodic boundary conditions on the sides of a rectangular domain of {\it 2D component}.
At the bottom of the rectangle, the external current density induced by the axion, corresponding to Eq.~\ref{eq:axion_haloscope_maxwell}, was driven with a boundary condition of a perfect electric conductor.
At the top, a perfect matched layer was placed to obtain the power radiated to the vacuum without reflection.
For verification, a region with a length of 63~mm (10.5 wavelengths) under a magnetic field of 10~T was prepared with a target frequency near 50~GHz.
The simulation was conducted by sweeping the possible axion frequencies with $g_{\gamma}=0.97$ and $\varrho_{a} = 0.45\,{\rm GeV/cm^{3}}$.
These model-dependent quantities only affect the magnitude of the conversion power.
Fig.~\ref{fig:haloscope_simul}~(a) shows the resulting conversion power obtained by the simulation and also by Eq.~\ref{eq:DAH_power}.
The conversion power is exactly given as expected and follows the sine-to-the-power-of-4 function with respect to the frequency.
Their differences are generally below the percent level.

\subsection{Cavity haloscope}
A microwave cavity coherently accumulates the photons converted from the axions inside the cavity when its resonant frequency matches the frequency of the dark matter axion.
This method, called a cavity haloscope, was proposed by P. Sikivie~\cite{PRD1985Sikivie}, and improves conversion power by a factor of quality of resonance.
The cavity haloscope is currently the most sensitive experimental method.
Since the axion frequency is {\it a priori} unknown, the resonant frequency of the cavity must be tuned and the possible axion frequencies are scanned.

The reacted field and the conversion power near the resonance is approximately given by~\cite{JCAP2020Kim}:
\begin{equation}
\label{eq:CAV_power}
\begin{split}
    \mathbf{E}_{r}  & \approx \mathbf{E}_{m} + \frac{g_{a\gamma\gamma}a_{0}}{\sqrt{\varepsilon \mu_{0}}} \mathbf{B}_{o}, \\
    P_{a\gamma\gamma} & \approx \left[ \frac{g_{a\gamma\gamma}^{2}\varrho_{a}}{m_{a}^{2}c/\hbar^{3}} \right] \left[\omega_{a} \int \frac{1}{\mu} \left|\mathbf{B}_{o}\right|^{2} dV_{c} \right] \times \\
    & \qquad C \frac{Q_{c}Q_{a}}{Q_{c} + Q_{a}}\frac{ 1 }{((2 Q_{c}(\omega_{a} - \omega_{c}) / \omega_{c})^{2} + 1},
\end{split}
\end{equation}
where $\mathbf{E}_{m}$ is the reacted electric field forming the cavity mode, $V_{c}$ is the cavity volume, $\omega_{c}$ is the cavity's angular resonant frequency, and $Q_{c}$ and $Q_{a}$ are the quality factors of cavity and axion, respectively.
$C$ is the form factor which represents how well the mode electric field is aligned with the applied magnetic field.
\begin{equation}
    C = \frac{\left| \int \mathbf{E}_{r} \cdot \mathbf{B}_{o} dV_{c} \right|^{2}}{\int \varepsilon_{r}|\mathbf{E}_{r}|^{2} dV_{c} \int |\mathbf{B}_{o}|^{2} dV_{c}},
\end{equation}
where $\varepsilon_{r}=\varepsilon / \varepsilon_{0}$ is the relative permittivity.

Similarly, for verification, the reacted field by the dark matter axion in an ideal cylindrical cavity was simulated with \texttt{COMSOL}'s {\it 2D axisymmetric component}.
$Q_{a}$ is defined as the distribution of the dark matter axion in the frequency domain.
Here, a monochromatic axion ($Q_{a}\to \infty$) is assumed for simulation convenience.
The conversion power was calculated for a cylindrical copper cavity with a radius of 50~mm and a height of 100~mm under a uniform magnetic field of 10~T applied along the cylinder axis.
The same values as in Section~\ref{subsec:dah} were used for the axion-photon coupling and the local dark matter density.
The simulation results were compared with the conversion power of the analytic solution, as shown in Fig.~\ref{fig:haloscope_simul}~(b).
Since the signal is enhanced with the cavity resonance, we simulated the axion conversion power around the resonant frequency of 2295~MHz determined by the radius of the cavity.
In this case too, the simulated power and the approximate power obtained through analytic calculation are almost identical.

\subsection{Dielectric haloscope}
Similar to metal surfaces, dark matter axions on dielectric surfaces with different dielectric constants are converted into photons under a magnetic field.
The phase of the photons converted at each dielectric surface is roughly the same within the axion's de Broglie wavelength.
When dielectrics are periodically arranged at intervals of the axion's Compton wavelength, the electromagnetic waves generated on each dielectric surface constructively interfere as they propagate, boosting the total conversion power.
This enables the dielectric haloscope to be effective for axion search at higher frequencies up to 50\,GHz~\cite{PRL2017MADMAX}.

In this scheme, an analytic solution has been studied for a finite length of an array of dielectric plates with infinite radius~\cite{JCAP2017Millar}.
The MADMAX group has also seen the effect of the finite radius of dielectric plates in a 3-dimensional space with a similar FEM-based method~\cite{JCAP2019Knirck}.
For verification purposes, we consider the simplest case of a single dielectric disk with an infinite radius.
When a dielectric of a relative permittivity $\varepsilon_{r}$ with thickness $d$ is placed at the position of $x = 0$, and a magnetic field of $B_{0}$ is applied with a length of $L$ on both sides. The reacted field solution propagating outward is as follows.
\begin{equation}
    \mathbf{E}_{r} = -i c g_{a\gamma\gamma}a_{0} B_{0} e^{i k x} e^{-\frac{i k d}{2}} \upbeta \hat{y}.
\end{equation}
Here the complex enhancement factor $\upbeta$ is
\begin{equation}
\begin{split}
    \upbeta &= \left[ \sqrt{\varepsilon_{r}}\cos\left( \frac{1}{2}d k \sqrt{\varepsilon_{r}}\right)\sin\left( k L \right) \right. \\
    & \left. + (1 - \varepsilon_{r} + \varepsilon_{r} \cos\left(k  L \right))\sin\left(\frac{1}{2}d k \sqrt{\varepsilon_{r}} \right) \right] \\
    & \Big/ \left[ \varepsilon_{r} \sin\left(\frac{1}{2}d k \sqrt{\varepsilon_{r}} \right) + i \sqrt{\varepsilon_{r}}\cos\left(\frac{1}{2} d k \sqrt{\varepsilon_{r}} \right) \right].
\end{split}
\end{equation}
The radiation power emitted in one direction is given from the Poynting vector of the field solution, as follows.
\begin{equation}
\label{eq:DEH_power}
    \frac{P_{a\gamma\gamma}}{A} = \left[ \frac{g_{a\gamma\gamma}^{2}\varrho_{a}}{m_{a}^{2}c /\hbar^{3}} \right] \frac{c}{\mu}|\mathbf{B}_{o} \times \hat{n}|^{2} \beta^{2},
\end{equation}
where the enhancement factor $\beta$ is
\begin{equation}
\begin{split}
    \beta^{2} &= 2 \left[ \sqrt{\varepsilon_{r}}\cos\left( \frac{1}{2}d k \sqrt{\varepsilon_{r}}\right)\sin\left( k L \right)\right.\\
    &\left. + (1 - \varepsilon_{r} + \varepsilon_{r} \cos\left(k  L \right))\sin\left(\frac{1}{2}d k \sqrt{\varepsilon_{r}} \right)\right]^{2} \\
    & \Big/ \left[ \varepsilon_{r} \left( (\varepsilon_{r} + 1) - (\varepsilon_{r} - 1) \cos\left(d k \sqrt{\varepsilon_{r}} \right) \right) \right].
\end{split}
\end{equation}

\begin{figure}
    \centering
    \includegraphics[width=\linewidth]{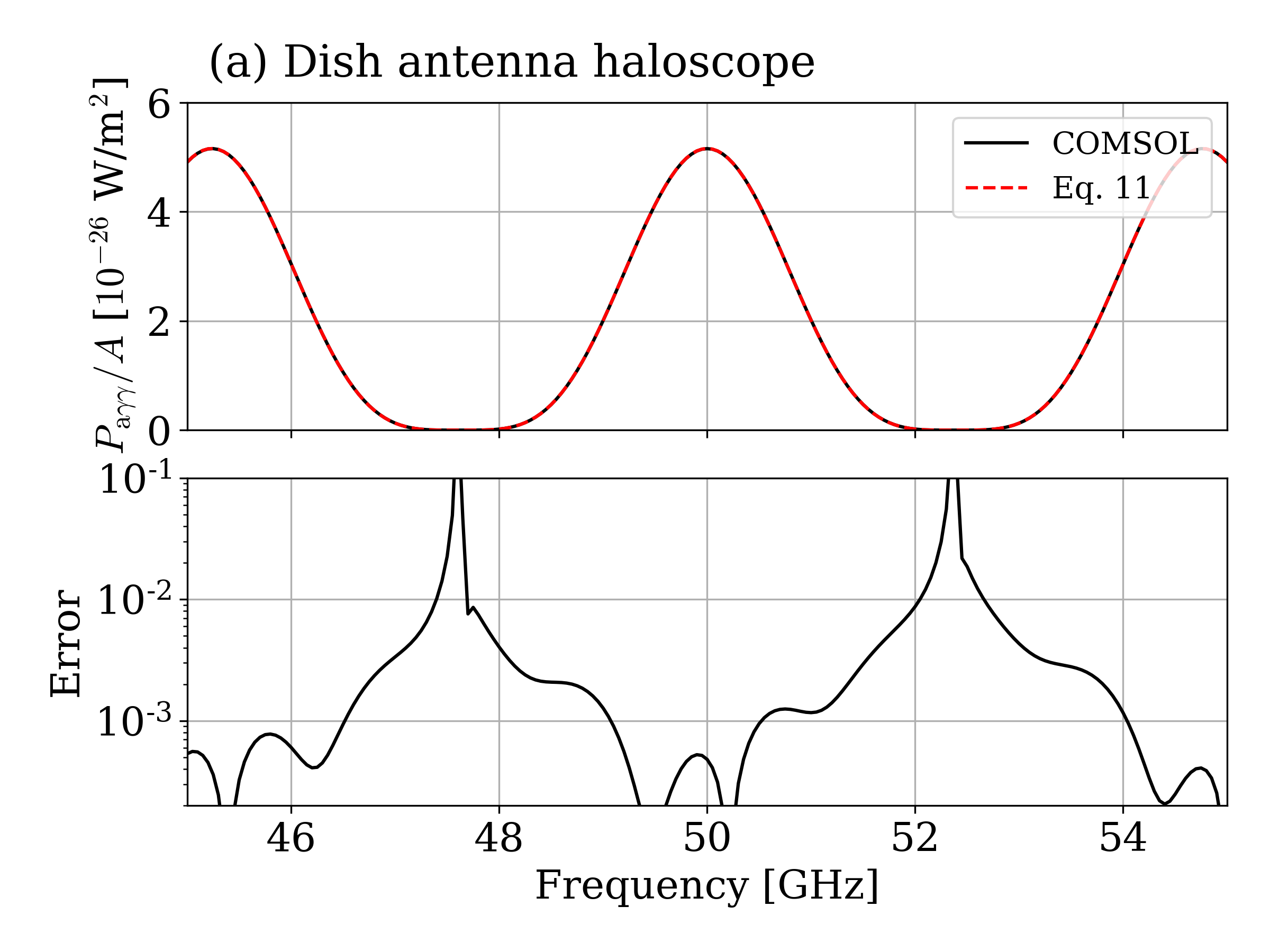}
    \includegraphics[width=\linewidth]{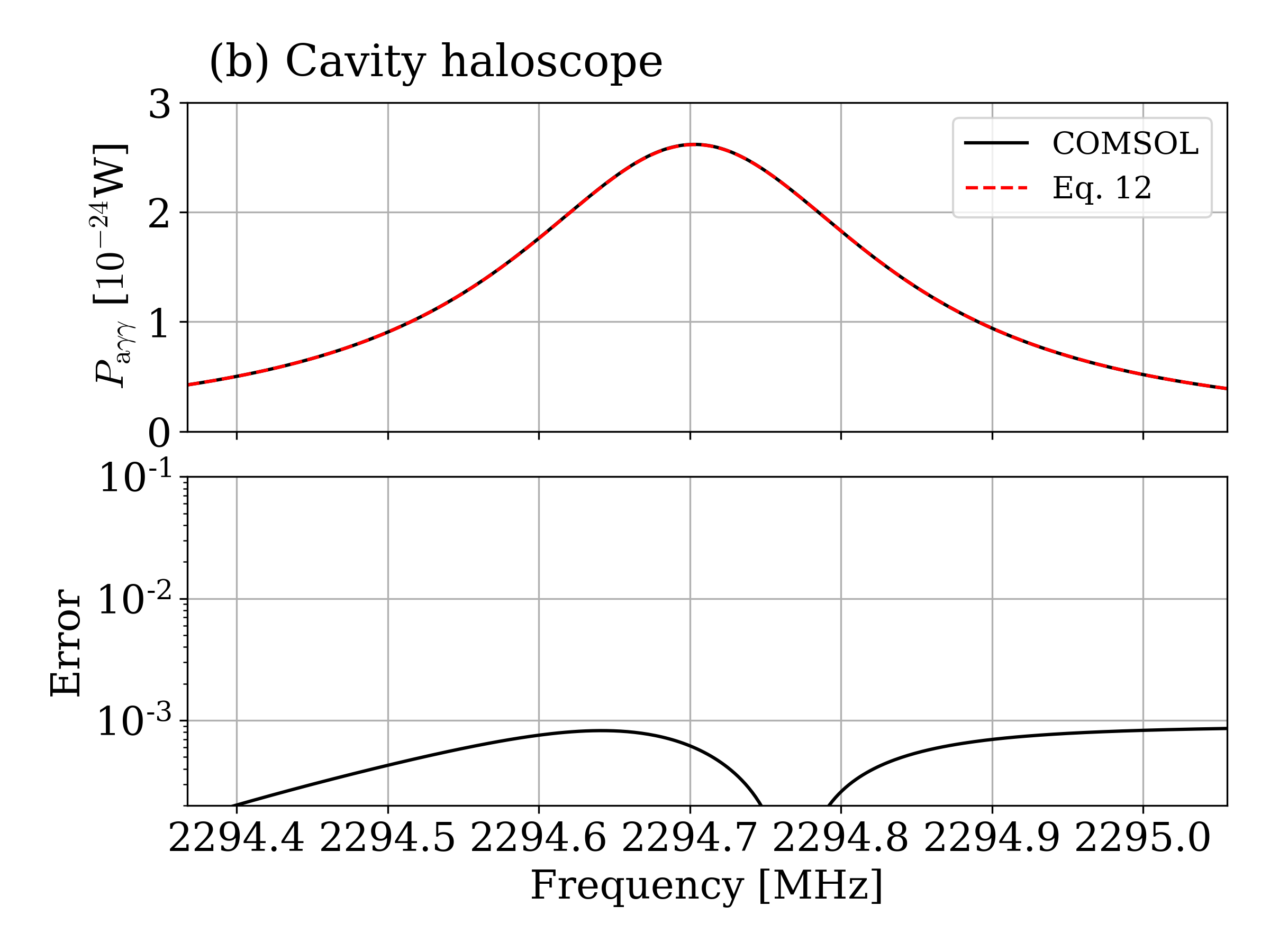}
    \includegraphics[width=\linewidth]{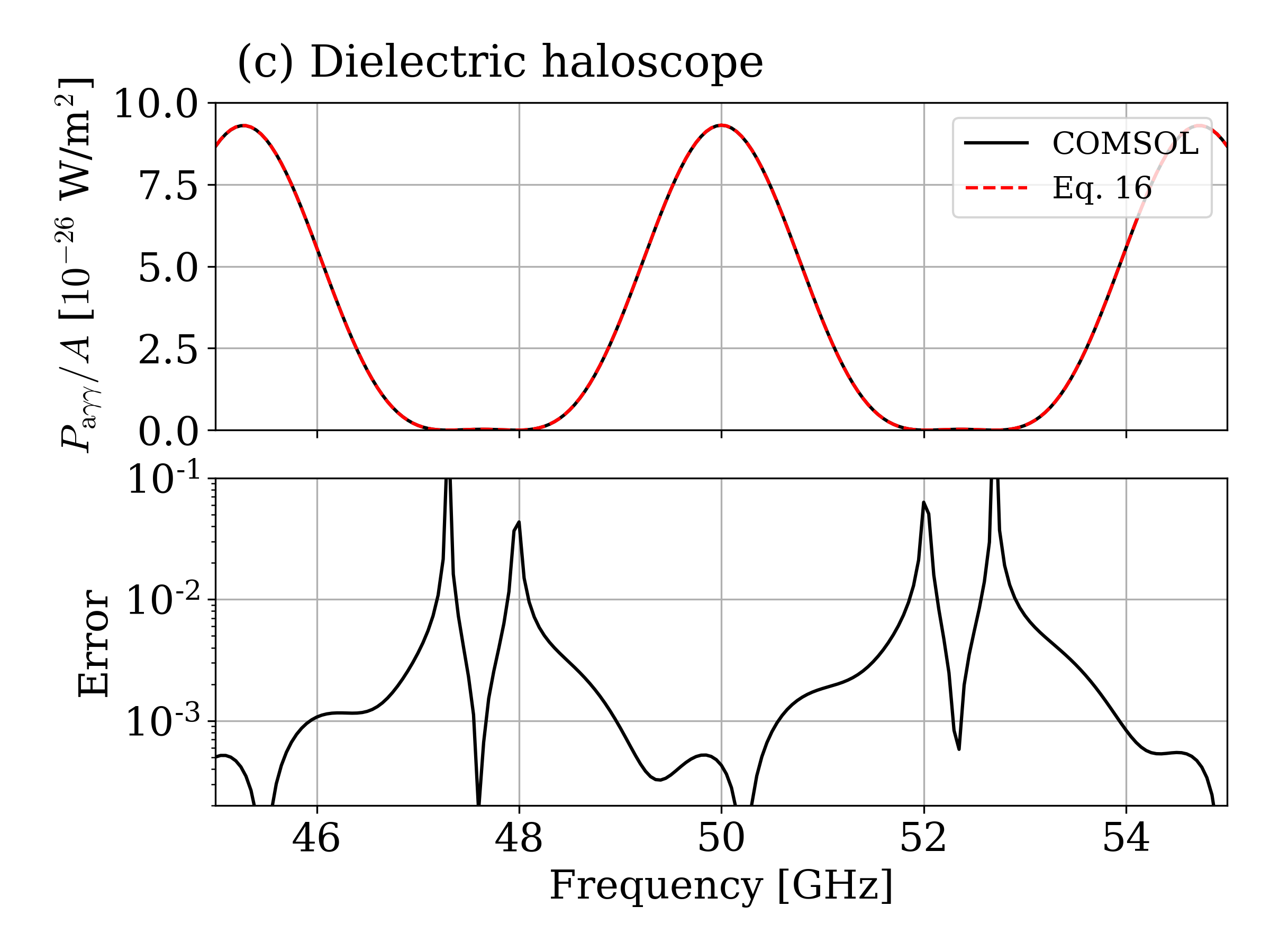}
    \caption{Comparison and relative difference of conversion power obtained through numerical simulation and analytical solution for (a) dish antenna haloscope, (b) cavity haloscope, and (c) dielectric haloscope. 
    The error is defined as the difference between the simulation and analytic solution normalized to the latter.
    The large variations in the error distribution is mostly attributed to the mesh configuration.
    A local dark matter composed solely by the KSVZ axion is assumed.
    The geometry of the individual haloscopes is described in the text.}
    \label{fig:haloscope_simul}
\end{figure}

As before, for comparison, the conversion power was calculated to be around 50~GHz when a 10~T magnetic field was applied to 63~mm ($L$) long regions on both sides of a 0.95~mm ($d$) thick dielectric with a relative permittivity of 10 ($\varepsilon_{r}$).
The conversion power obtained from the analytic solution and the power obtained from \texttt{COMSOL} are compared in Fig.~\ref{fig:haloscope_simul}~(c).
It was confirmed that the overall error was below the percent level.

\section{\label{sec:axion2nd}Axion-Photon Regeneration}
There are axion search experiments that do not assume dark matter.
These experiments attempt to observe second-order effects due to the creation and annihilation of the axion fields.
It is less sensitive than haloscope experiments because the reacted electromagnetic field being observed is proportional to the square of the axion-to-photon coupling.

These experiments can be simulated by division into three processes: (1) excitation and (2) propagation of the axion, and (3) regeneration of the photon.
The excitation and propagation of the axion field use the Klein-Gordon equation.
\begin{equation}
\label{eq:axion_eom}
    \left(\frac{1}{c^2} \partial_{t}^{2} - \nabla^{2} + \frac{m_{a}^{2} c^{2}}{\hbar^{2}} \right) a = -g_{a\gamma\gamma} \frac{\hbar c}{Z_{0}} \mathbf{E}\cdot \mathbf{B}.
\end{equation}
The process of photon regeneration employs the modified Maxwell's equations as described in the previous section.
\texttt{COMSOL} can also apply the Klein-Gordon equation with the physics of \textit{Coefficient Form PDE} to simulate all these processes.

One of the representative experiments is Axion-Photon Regeneration~\cite{PRL1987Bibber,PLB2010Ehret,PRD2015Ballou,PRL2007Sikivie}.
Photons propagating under a strong magnetic field induce an axion field by Eq.~\ref{eq:axion_eom}.
The induced axion field propagates freely through space with the same energy as the injected photon.
Axions can easily cross an obstacle that photons cannot pass.
That is, photons injected from one side of the obstacle are blocked, but photons of the same frequency from the other side of the obstacle can be generated by the produced axions under a magnetic field.

When a magnetic field $B_{0}$ is applied to a cavity of length $L$, the conversion probability from the incident photon to the axion is calculated as follows~\cite{PRL2007Sikivie}.
\begin{equation}
\label{eq:APR_prob}
    p \approx \left( \frac{2 k_{\gamma}}{k_{\gamma} + k_{a}} \right)^{2} \left(g_{a\gamma\gamma}B_{0}\sqrt{\frac{\hbar c}{\mu_{0}}} \right)^{2} \frac{Q_{c} L}{4 k_{a}} F( \left| k_{\gamma} - k_{a} \right| ) ,
\end{equation}
where $k_{\gamma}=\omega/c$ is the incident photon's momentum, $k_{a} = \sqrt{k_{\gamma}^{2} - m_{a}^{2}c^{2}/\hbar^{2}}$ is the converted axion's momentum, $Q_{c}$ is the cavity quality factor, and $F$ is the form factor with the definition of:
\begin{equation}
    F(q) = \left[\frac{2}{q L} \sin\left(\frac{q L}{2} \right) \right]^{2}
\end{equation}
The first parenthetical term in Eq.~\ref{eq:APR_prob} is an additional factor due to the momentum transfer from the inhomogeneous magnetic field.
When the cavity is not sufficiently longer than the wavelength of the converted axion, the factor affects the probability.
In an actual experimental setup, since cavities are sufficiently long compared to the photon wavelength, it can be neglected.

\begin{figure}
    \centering
    \includegraphics[width=\linewidth]{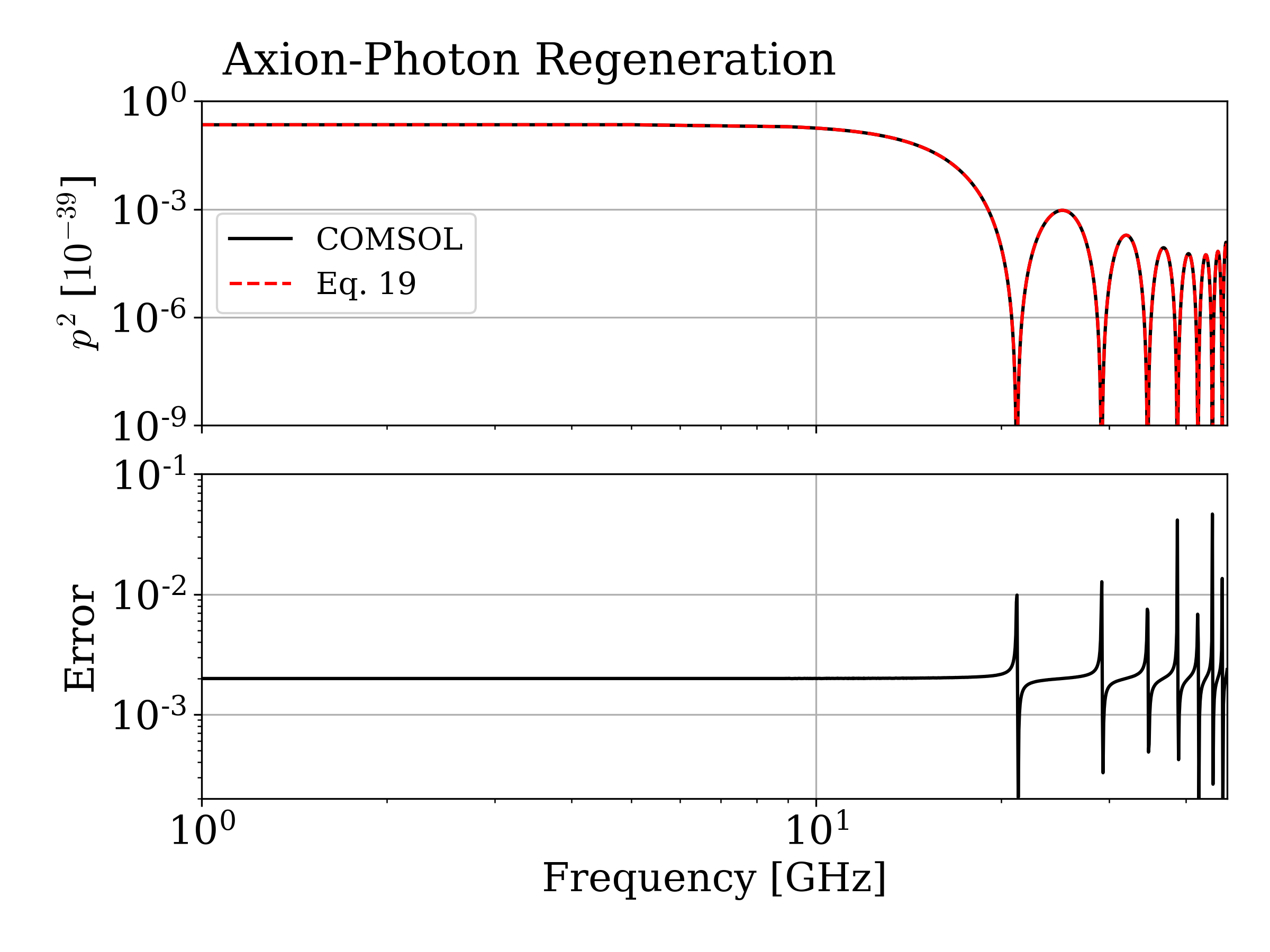}
    \caption{Comparison and relative difference of photon energy ratio obtained through numerical simulation and analytical solution between the two cavities for the photon regeneration scheme depicted in the text.}
    \label{fig:APR_probability}
\end{figure}

For verification, the ratio of photon energy transferred to the opposite side by the axion was calculated.
50~GHz photons were injected into one cavity with a length ($L$) of 60~mm and the other cavity is located behind an obstacle.
Similar to the previous calculations, an infinite cross-section cavity was assumed by applying periodic boundary conditions to  both sides of a rectangle domain in \texttt{COMSOL}'s {\it 2D component}.
For sufficient conversion, it was calculated for a $g_{a\gamma\gamma}$ of $10^{-11}$~GeV$^{-1}$ in a magnetic field of 10~T.
The cavity had a quality factor of about $10^{5}$ assuming copper conductivity for both ends of the cavity.
The conversion probability was calculated as the ratio of energies stored in each cavity, and plotted along with the analytic solution in Fig.~\ref{fig:APR_probability}.
Similarly, it was confirmed that the results of \texttt{COMSOL} and the analytic solution were almost identical at the sub percent level.

\section{\label{sec:summary}Summary}
In this work, we reviewed the classical electrodynamics including the axion-photon interaction and demonstrated that they can be numerically simulated using a commercially available software, \texttt{COMSOL Multiphysics}.
With the axion-induced source terms implemented in the simulation, the reacted electromagnetic fields were numerically computed for various haloscope schemes.
For verification, the conversion powers were calculated and compared with the analytical results.
We also showed that axion generation and propagation can be simulated by considering the Klein-Gordon equation.
This study suggests that a numerical approach will help design and optimize future axion search experiments.

\begin{acknowledgments}
This work was supported by the Institute for Basic Science (IBS-R017-D1-2023-a00).
\end{acknowledgments}


\begin{thebibliography}{0}%
\makeatletter
\providecommand \@ifxundefined [1]{%
 \@ifx{#1\undefined}
}%
\providecommand \@ifnum [1]{%
 \ifnum #1\expandafter \@firstoftwo
 \else \expandafter \@secondoftwo
 \fi
}%
\providecommand \@ifx [1]{%
 \ifx #1\expandafter \@firstoftwo
 \else \expandafter \@secondoftwo
 \fi
}%
\providecommand \natexlab [1]{#1}%
\providecommand \enquote  [1]{``#1''}%
\providecommand \bibnamefont  [1]{#1}%
\providecommand \bibfnamefont [1]{#1}%
\providecommand \citenamefont [1]{#1}%
\providecommand \href@noop [0]{\@secondoftwo}%
\providecommand \href [0]{\begingroup \@sanitize@url \@href}%
\providecommand \@href[1]{\@@startlink{#1}\@@href}%
\providecommand \@@href[1]{\endgroup#1\@@endlink}%
\providecommand \@sanitize@url [0]{\catcode `\\12\catcode `\$12\catcode
  `\&12\catcode `\#12\catcode `\^12\catcode `\_12\catcode `\%12\relax}%
\providecommand \@@startlink[1]{}%
\providecommand \@@endlink[0]{}%
\providecommand \url  [0]{\begingroup\@sanitize@url \@url }%
\providecommand \@url [1]{\endgroup\@href {#1}{\urlprefix }}%
\providecommand \urlprefix  [0]{URL }%
\providecommand \Eprint [0]{\href }%
\providecommand \doibase [0]{http://dx.doi.org/}%
\providecommand \selectlanguage [0]{\@gobble}%
\providecommand \bibinfo  [0]{\@secondoftwo}%
\providecommand \bibfield  [0]{\@secondoftwo}%
\providecommand \translation [1]{[#1]}%
\providecommand \BibitemOpen [0]{}%
\providecommand \bibitemStop [0]{}%
\providecommand \bibitemNoStop [0]{.\EOS\space}%
\providecommand \EOS [0]{\spacefactor3000\relax}%
\providecommand \BibitemShut  [1]{\csname bibitem#1\endcsname}%
\let\auto@bib@innerbib\@empty
\end{thebibliography}%


\begin{references}
\bibitem{PRL1977PQ} R. D. Peccei and H. R. Quinn, Phys. Rev. Lett. {\bf 38}, 1440 (1977).
\bibitem{PRL1978Weinberg} S. Weinberg, Phys. Rev. Lett. {\bf 40}, 223 (1978).
\bibitem{PRL1978Wilczek} F. Wilczek, Phys. Rev. Lett. {\bf 40}, 279 (1978).
\bibitem{PLB1983Wilczek} J. Preskill, M. B. Wise and F. Wilczek, Phys. Lett. B {\bf 120}, 127 (1983).
\bibitem{PLB1983Abbott} L.F. Abbott and P. Sikivie, Phys. Lett. B {\bf 120}, 133 (1983).
\bibitem{PLB1983Dine} M. Dine and W. Fischler, Phys. Lett. B {\bf 120}, 137 (1983).
\bibitem{PRL1979Kim} J. E. Kim, Phys. Rev. Lett. {\bf 43}, 103 (1979).
\bibitem{NPB1980SVZ} M. A. Shifman, A. I. Vainshtein and V. I. Zakharov, Nucl. Phys. B {\bf 166}, 493 (1980).
\bibitem{YF1980Zhitnitsky} A. P. Zhitnitsky, Yad. Fiz. {\bf 31}, 497 (1980); Sov. J. Nucl. Phys. {\bf 31} (1980).
\bibitem{PLB1981DFS} M. Dine, W. Fischler and M. Srednicki, Phys. Lett. B {\bf 104}, 199 (1981).
\bibitem{PRL1987Wilczek} F. Wilczek, Phys. Rev. Lett. {\bf 58}, 1799 (1987).
\bibitem{PRD1985Sikivie} P. Sikivie, Phys. Rev. D {\bf 32}, 2988 (1985).
\bibitem{PRL2017MADMAX} A. Caldwell {\it et al.} (MADMAX Working Group), Phys. Rev. Lett. {\bf 118}, 091801 (2017).
\bibitem{JCAP2013Horns} D. Horns {\it et al.}, J. Cosmol. Astropart. Phys. 04 (2013) 016.
\bibitem{PRL1987Bibber} K. A. van Bibber {\it et al.}, Phys. Rev. Lett. {\bf 59}, 759 (1987).
\bibitem{COMSOL} COMSOL Multiphysics$\textsuperscript{\textregistered}$ v. 5.2. www.comsol.com. COMSOL AB, Stockholm, Sweden.
\bibitem{JPG2014Read} J. I. Read, J. Phys. G: Nucl. Part. Phys. {\bf 41}, 063101 (2014).
\bibitem{NATURE2017Borsanyi} S. Borsanyi {\it et al.}, Nature {\bf 539}, 69 (2016).
\bibitem{PDU2019Kim} Y. Kim {\it et al.}, Phys. Dark Universe {\bf 26}, 100362 (2019).
\bibitem{arxiv2023DMRadio} A. AlShirawi {\it et al.} (DMRadio Collaboration), arXiv:2302.14084 (2023).
\bibitem{JCAP2020Kim} D. Kim {\it et al.}, J. Cosmol. Astropart. Phys. 03 (2020) 066.
\bibitem{JCAP2017Millar} A. J. Millar {\it et al.}, J. Cosmol. Astropart. Phys. 01 (2017) 061.
\bibitem{JCAP2019Knirck} S. Knirck {\it el al.}, J. Cosmol. Astropart. Phys. 08 (2019) 026.
\bibitem{PLB2010Ehret} K. Ehret {\it et al.}, Phys. Lett. B {\bf 689}, 149 (2010).
\bibitem{PRD2015Ballou} R. Ballou {\it et al.}, Phys. Rev. D {\bf 92}, 092002 (2015).
\bibitem{PRL2007Sikivie} P. Sikivie, D. B. Tanner, and K. A. van Bibber, Phys. Rev. Lett. {\bf 98}, 172002 (2007).

\end{references}
\end{document}